\documentstyle[psfig,draft]{mn}

\newif\ifAMStwofonts

\newcommand{\uflux}{{\rm erg}\, {\rm cm}^{-2}\, {\rm s}^{-1}}
\newcommand{\uint}{{\rm keV}\, {\rm cm}^{-2}\, {\rm s}^{-1}\, {\rm sr}^{-1}\, {\rm keV}^{-1}}
\newcommand{\keV}{{\rm keV}}
\newcommand{\ASCA}{{\it ASCA}}
\newcommand{\SAX}{{\it BeppoSAX}}
\newcommand{\ROSAT}{{\it ROSAT}}

\ifoldfss
  \ifCUPmtlplainloaded \else
    \NewTextAlphabet{textbfit} {cmbxti10} {}
    \NewTextAlphabet{textbfss} {cmssbx10} {}
    \NewMathAlphabet{mathbfit} {cmbxti10} {} 
    \NewMathAlphabet{mathbfss} {cmssbx10} {} 
  \fi
  \ifAMStwofonts
    \ifCUPmtlplainloaded \else
      \NewSymbolFont{upmath} {eurm10}
      \NewSymbolFont{AMSa} {msam10}
      \NewMathSymbol{\upi}     {0}{upmath}{19}
      \NewMathSymbol{\umu}     {0}{upmath}{16}
      \NewMathSymbol{\upartial}{0}{upmath}{40}
      \NewMathSymbol{\leqslant}{3}{AMSa}{36}
      \NewMathSymbol{\geqslant}{3}{AMSa}{3E}

    \fi
  \fi
\fi 

\ifnfssone
  \newmathalphabet{\mathit}
  \addtoversion{normal}{\mathit}{cmr}{m}{it}
  \addtoversion{bold}{\mathit}{cmr}{bx}{it}
  \newmathalphabet{\mathbfit} 
  \addtoversion{normal}{\mathbfit}{cmr}{bx}{it}
  \addtoversion{bold}{\mathbfit}{cmr}{bx}{it}
  \newmathalphabet{\mathbfss} 
  \addtoversion{normal}{\mathbfss}{cmss}{bx}{n}
  \addtoversion{bold}{\mathbfss}{cmss}{bx}{n}
  \ifAMStwofonts
    \ifCUPmtlplainloaded \else
      \UseAMStwoboldmath
      \makeatletter
      \new@mathgroup\upmath@group
      \define@mathgroup\mv@normal\upmath@group{eur}{m}{n}
      \define@mathgroup\mv@bold\upmath@group{eur}{b}{n}
      \edef\UPM{\hexnumber\upmath@group}
      \new@mathgroup\amsa@group
      \define@mathgroup\mv@normal\amsa@group{msa}{m}{n}
      \define@mathgroup\mv@bold\amsa@group{msa}{m}{n}
      \edef\AMSa{\hexnumber\amsa@group}
      \makeatother
      \mathchardef\upi="0\UPM19
      \mathchardef\umu="0\UPM16
      \mathchardef\upartial="0\UPM40
      \mathchardef\leqslant="3\AMSa36
      \mathchardef\geqslant="3\AMSa3E
    \fi
  \fi
\fi 

\ifnfsstwo
  \DeclareMathAlphabet{\mathbfit}{OT1}{cmr}{bx}{it}
  \SetMathAlphabet\mathbfit{bold}{OT1}{cmr}{bx}{it}
  \DeclareMathAlphabet{\mathbfss}{OT1}{cmss}{bx}{n}
  \SetMathAlphabet\mathbfss{bold}{OT1}{cmss}{bx}{n}
  \ifAMStwofonts
    \ifCUPmtlplainloaded \else
      \DeclareSymbolFont{UPM}{U}{eur}{m}{n}
      \SetSymbolFont{UPM}{bold}{U}{eur}{b}{n}
      \DeclareSymbolFont{AMSa}{U}{msa}{m}{n}
      \DeclareMathSymbol{\upi}{0}{UPM}{"19}
      \DeclareMathSymbol{\umu}{0}{UPM}{"16}
      \DeclareMathSymbol{\upartial}{0}{UPM}{"40}
      \DeclareMathSymbol{\leqslant}{3}{AMSa}{"36}
      \DeclareMathSymbol{\geqslant}{3}{AMSa}{"3E}
    \fi
  \fi
\fi 

\ifCUPmtlplainloaded \else
  \ifAMStwofonts \else 
    \def\upi{\pi}
    \def\umu{\mu}
    \def\upartial{\partial}
  \fi
\fi

\title{On the intensity of the extragalactic X-ray background}
\author[X. Barcons et al.]
       {X. Barcons$^1$, S. Mateos$^{1,2}$ M.T. Ceballos$^1$ \\
        $^1$ Instituto de F\'\i sica de Cantabria (CSIC-UC), 39005 Santander, Spain\\
	$^2$ Departamento de F\'\i sica Moderna, Universidad de Cantabria, 39005 Santander, Spain\\
}
\date{May 2000}

\pagerange{\pageref{firstpage}--\pageref{lastpage}}
\pubyear{2000}

\begin{document}

\maketitle

\label{firstpage}

\begin{abstract}
Measurements of the intensity of the cosmic X-ray background (XRB)
carried out over small solid angles are subject to spatial variations
due to the discrete nature of the XRB. This cosmic variance can
account for the dispersion of XRB intensity values found within the
\ASCA, \SAX\ and \ROSAT\ missions separately. However there are
differences among the values obtained in the different missions which
are not due to spatial fluctuations but, more likely, to systematic
cross-calibration errors. Prompted by recent work which shows that
\ROSAT\ PSPC has calibration differences with all the other missions,
we compute a bayesian estimate for the XRB intensity at 1 keV of
$10.0^{+0.6}_{-0.9}\, \uint$ (90 per cent confidence errors) using the
\ASCA\ and \SAX\ data points. However, this value is still
significantly larger than the $HEAO-1$ intensity measured over many
thousands of square degrees ($8\, \uint$).
\end{abstract}


\section{Introduction}

Most of the X-ray background (XRB) above photon energies of a few keV
is known to be extragalactic in origin (see, e.g, reviews by Boldt
1987 and Fabian \& Barcons 1992). Its spectrum was well measured by
the $HEAO-1$ A2 experiment in the 3-50 keV energy range (Marshall et
al 1980), where it fits a thermal bremsstrahlung model with a
temperature $kT\approx 40\, \keV$.  An overall fit to the XRB spectrum
from 3 keV to 10 MeV was presented by Gruber (1992) based on data from
the A2 and A4 experiments on board $HEAO-1$, which showed an
extrapolated intensity at 1 keV of $\approx 8 \uint$.

More recently, measurements of the XRB spectrum and intensity at
photon energies $< 10\, \keV$ have been obtained with the use of imaging
instruments. Using \ASCA\ data, Gendreau et al (1995) confirmed that
the thermal bremsstrahlung shape (which at photon energies below $\sim
20\, \keV$ can be approximated by a power law with energy spectral
index $\alpha\sim 0.37$) provides a good fit to the XRB spectrum
down to about 1 keV.  Vecchi et al (1999) used the LECS and MECS
instruments on board \SAX\ in the 1-8 keV band to confirm that the XRB
spectrum is consistent with a power law with $\alpha\sim 0.4$.

$ROSAT$ PSPC observations have also provided data on the spectrum of
the extragalactic XRB.  Hasinger (1992) found a very steep spectrum
for the XRB below 2 keV ($\alpha\sim 1$).  However, shadowing
experiments with $ROSAT$ have provided the more stringent upper limits
to the slope of the extragalactic XRB at soft X-ray energies
($\alpha<0.7$, Barber \& Warwick 1994). The relative steepness of the
\ROSAT\ spectra with respect to many other instruments has been
studied in detail by Iwasawa, Fabian \& Nandra (1999). The fact that
the spectral shape of the same sources is usually consistent in
observations of many instruments with the exception of \ROSAT\ (which
usually finds steeper spectra) is highly suggestive of a calibration
mismatch which will certainly affect the XRB spectrum as well.

The normalisation of the extragalactic XRB (that we parametrise as the
XRB intensity $I_{XRB}$ at 1 keV in units of $\uint$) remains
uncertain.  The Marshall et al (1980) measurement is the most robust
result as it was performed over a very large solid angle ($\sim 10^4\,
\deg^2$) with instrumentation especially designed to subtract
efficiently and accurately the particle background.  The XRB intensity
measured by the imaging instruments on board \ASCA, \SAX\ and \ROSAT\
invariably yield higher values which are often statistically
discrepant among them.

\begin{table*}
\begin{center}
\begin{tabular}{l l l c c c l}
\hline
Instrument & Field & $\Omega_i$ & $I_i$ & $\sigma_i$ & $\Sigma_i$ & Ref\\
           & Name  & ($\deg^2$) & ($\uint$)\\ 
\hline
ROSAT PSPC & QSF3  & 0.223      & 11.4  &  0.34      &  1.00       &
Chen et al (1996)\\
ASCA SIS   & QSF3  & 0.134      & 10.0  &  0.37      &  0.76       &
Chen et al (1996)\\
ROSAT PSPC & GSGP4 & 0.283      & 11.8  &  0.44      &  4.49      &
Georgantopoulos et al (1996)\\
ROSAT PSPC & SGP2  & 0.283      & 12.0  &  0.61      &  4.53      & 
Georgantopoulos et al (1996)\\
ROSAT PSPC & SGP3  & 0.283      & 12.0  &  0.68      &  4.52      &
Georgantopoulos et al (1996)\\
ROSAT PSPC & QSF1  & 0.283      & 9.9   &  0.65      &  4.49      &
Georgantopoulos et al (1996)\\
ASCA GIS   & Lockman    & 0.165      & 10.9  &  0.61      &  1.31      &
Miyaji et al (1998)\\
ASCA GIS   & Lynx    & 0.144      &  9.3  &  0.61      &     1.38   &
Miyaji et al (1998)\\
ROSAT PSPC & Lockman    & 0.165      & 10.0  &  0.30      &   2.10     &
Miyaji et al (1998)\\
ROSAT PSPC & Lynx    & 0.144      & 11.5  &  0.43      &    2.28     &
Miyaji et al (1998)\\
ASCA SIS   & Various& 0.538     &  8.9  &  0.50      &     0.96   &
Gendreau et al (1995)\\
ROSAT PSPC & Various& 5.9       & 13.4  &  0.18      &   0.66     &
Hasinger (1992)\\
SAX LECS+MECS  & Various& 0.726     & 11.0  &  0.30      &   0.77      &
Vecchi et al (1999)\\
\hline
\end{tabular}
\caption{Measurements of the XRB intensity at 1 keV carried out with
imaging instruments.$\Omega_i$ is the solid angle covered, $I_i$ the
measured intensity, $\sigma_i$ the 1-sigma statistical uncertainty and 
$\Sigma_i$ the 1-sigma uncertainty derived when statistical and cosmic 
variances are included.}
\end{center}
\end{table*}

In this paper we point out that the discrete nature of the XRB
introduces a cosmic variance in its intensity which is large for
observations carried out over small solid angles. This cosmic variance
is just the confusion noise caused by unresolved or non removed
sources in the images (Scheuer 1974, Barcons 1992). We find variations
of the order of 10 per cent for solid angles under 1$\deg^2$ and
therefore cosmic variance often dominates over the statistical
uncertainties quoted in the various estimates of the XRB intensity.
Once this is included, we combine individual measurements of the XRB
to compute bayesian estimates of the overall XRB intensity. We find
that measurements carried out within the same mission are brought to
consistency by cosmic variance, but that systematic differences
between different missions remain.

\section{The cosmic variance}

Table 1 lists the XRB intensity measurements that we have used. In
particular and for each particular datum we list the measured
intensity at 1 keV $I_i$, the quoted statistical 1-sigma error
$\sigma_i$ (converted from 90 per cent errors) and the solid angle
covered $\Omega_i$. It is clear that the variations between different
measurements carried out within the same mission are often
significant, if the statistical errors $\sigma_i$ are used to assess
the significance. In what follows we compute the additional uncertainty
introduced by the cosmic variance.

If the true average XRB intensity is $I_{XRB}$ then we want to compute
the probability density function $P_i(I|I_{XRB})$ of obtaining an
intensity $I$ in a measurement over a solid angle $\Omega_i$ with a
statistical error $\sigma_i$.  This is computed by first convolving
the confusion noise distribution with the statistical noise
distribution (which we assume to be gaussian) and later shifting the
resulting distribution to have a mean value $I_{XRB}$. This shift can
be attributed to the behaviour of the source counts in the very faint
regime, which would not affect the shape of the distribution. Assuming
a top-hat beam (which is appropriate for our purposes), the confusion
noise distribution can be expressed as (Barcons 1992)

\begin{equation}
P^{conf}_i(I)=\int d\omega\, {\rm e}^{-2\pi i\omega I}\exp(\Omega_i\int dS\,
N(S) [{\rm e}^{2\pi i\omega S}-1])
\end{equation}
where $N(S)$ is the number of sources per unit solid angle and unit
flux which in the range were we shall use it ($S\sim
10^{-13}-10^{-12}\, \uflux$) can be approximated as euclidean

\begin{equation} 
N(S)= {3\over 2} {K\over S_0} \left({S\over S_0}\right)^{-5/2},
\end{equation} 
where we choose $S_0=10^{-14}\, \uflux$ and then $K$ is the source
density, extrapolated with the same euclidean law, at a flux limit
$10^{-14}\, \uflux$. Indeed, the value of $K$ depends on the energy
band used for the observations. For the `hard' X-ray measurements
(\ASCA\ and \SAX ) we assume an extrapolation of the Piccinotti et al
(1982) source counts, now confirmed down to fluxes below $10^{-13}\,
\uflux$ (Cagnoni, Della Ceca \& Maccacaro 1998, Gendreau et al 1998).
This gives $K\approx 300\, \deg^{-2}$. For the `soft' X-ray
measurements we use the \ROSAT\ normalisation $K\approx 150\,
\deg^{-2}$ (Hasinger et al 1993, Branduardi-Raymont et al 1994). In
modelling the confusion noise distribution for the various
measurements of the XRB intensity, we have taken into account the
possible excision of bright foreground sources by applying the
corresponding cutoff in the flux integral of eqn. (1).  This really
only applies to the Chen et al (1996) \ROSAT\ measurement where
sources brighter than $3.4\times 10^{-14}\, \uflux$ were excluded.
For the remaining measurements we have assumed that the sources in the
Piccinotti et al (1982) catalogue had been avoided and therefore we
applied a cutoff at a flux of 2-10 keV of $3\times 10^{-11}\, \uflux$.

\begin{figure}
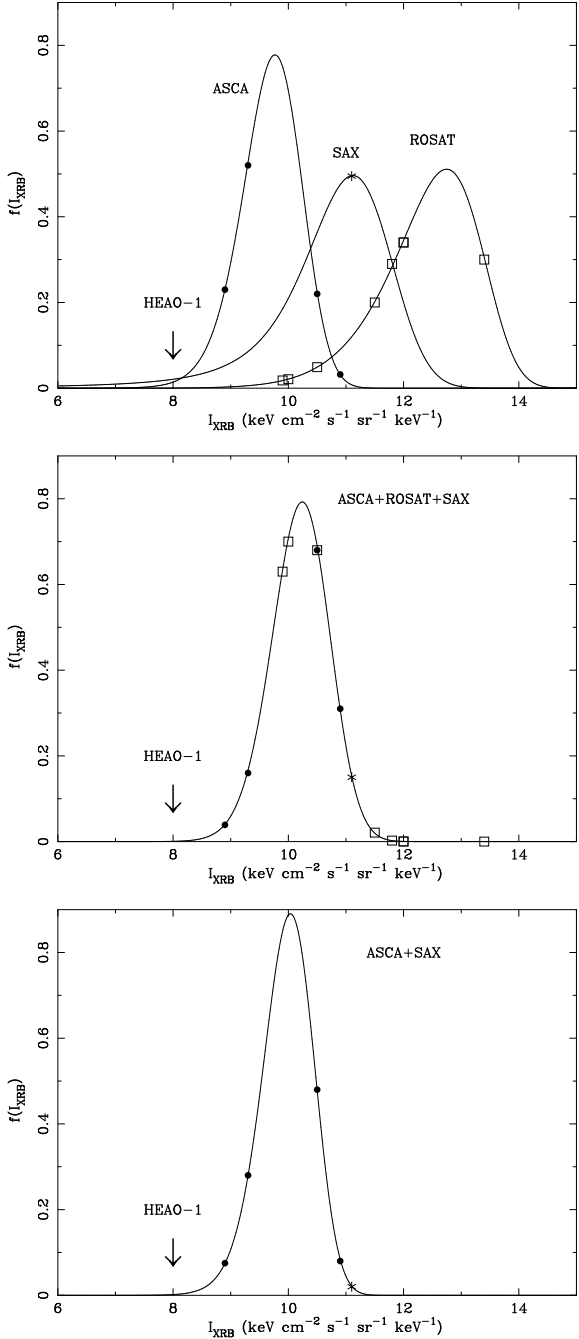

\psfig{file=fig1a.ps,height=5.7cm,angle=270.}
\vskip 0.3cm
\psfig{file=fig1b.ps,height=5.7cm,angle=270.}
\vskip 0.3cm
\psfig{file=fig1c.ps,height=5.7cm,angle=270.}
\caption{Bayesian density functions of the XRB intensity for the
measurements obtained with \ASCA, \SAX\ and \ROSAT . 
Individual measurements are shown on the same curves
as filled points (\ASCA ), asterisks (\SAX ) and squares (\ROSAT ).
The HEAO-1 point is also shown as a vertical arrow. The top panel shows the
bayesian distributions for each mission separately, the mid panel for the 3 missions all together and the bottom panel for the \ASCA\ and \SAX\ missions together.}
\end{figure}

Equation (1) ignores clustering of the XRB sources. Indeed clustering broadens the distribution
of intensities with respect to a uniform distribution (Barcons 1992).  From Barcons,
Fabian \& Carrera (1998) we see that for angular scales of the order of $\sim 1\deg^2$, excess fluctuations due to clustering amount to $\sim 0.5-1.2$\%, which implies a correction of the order of $\sim 10\%$ (at most) to the broadening computed by eq. (1). The relative amplitude of the clustering correction stays approximately constant when going to slightly smaller scales and decreases slowly when the small-scale declining part of the power spectrum of the fluctuations is reached.  Given the smallness of this effect end the uncertainties in the
modelling of the source clustering and its evolution we decided to ignore this
small effect.

Convolving eqn. 1 with a gaussian of dispersion $\sigma_i$ and
shifting the mean to a value $I_{XRB}$ gives the function
$P_i(I|I_{XRB})$ where now we express $I$ and $I_{XRB}$ as
monochromatic intensities at 1 keV in units of $\uint$.  For the
conversion between monochromatic intensity and broad-band flux $S$ per
unit solid angle in a given energy band we use an XRB spectral shape
with $\alpha=0.4$ for the `hard' instruments and $\alpha=0.7$ for the
`soft' instruments which result in

\begin{equation}
I(\uint)={S\over S_0}\, \Omega(\deg^2)^{-1}\, \Delta^{-1}
\end{equation}
where $\Delta\approx 200$ if the flux is in the 2-10 keV band and
$\Delta\approx 70$ if the flux is in the 0.5-2 keV band.

Typically, the curves including the cosmic variance are much broader
than the the statistical error. To illustrate this,
Table 1 also lists the dispersion $\Sigma_{i}$ obtained via a gaussian
fit to $P_i(I|I_{XRB})$ (which accounts for both the statistical and the
cosmic variance) for comparison with the statistical dispersion
$\sigma_i$. 

We now use these distributions to compute bayesian estimates of the
true XRB intensity $I_{XRB}$.  We assume an {\it a priori} distribution
which is uniform over a sufficiently wide range of values of $I_{XRB}$
(5 to 16 $\uint$).  The probability density function for $I_{XRB}$
given a set of data is (Press 1989)

\begin{equation}
f(I_{XRB})={\Pi_i\, P_i(I_i|I_{XRB})\over \int\, dI'_{XRB} \Pi_i\,
P_i(I_i|I'_{XRB})}
\end{equation}
where the products extend to the data points under consideration. In
what follows, our $I_{XRB}$ estimates correspond to the maximum of
that function, which we also use to derive 90\% confidence intervals.

\section{The intensity of the XRB}

We first compute $f(I_{XRB})$ for the measurements obtained within each imaging
X-ray mission separately. The results are shown in Fig. 1 where it can
be seen that the various measurements with the same instrument are
consistent when cosmic variance is taken into account.  In particular
the variety of \ASCA\ values which have been obtained with both the
SIS and GIS instruments at different epochs, does not call for extra
systematic effects.  The same comment applies to the \ROSAT\ data
points, as none of them is completely out of the distribution expected
in terms of cosmic variance and statistical uncertainties, in spite of
the fact that these intensities have been obtained with different
PSPCs and/or different gains. The corresponding estimates of the XRB
intensity are $9.8^{+0.7}_{-1.0}$, $11.1^{+1.0}_{-2.3}$ and
$12.7^{+0.9}_{-1.9}\, \uint$ (90 per cent errors) for \ASCA, \SAX\ and
\ROSAT\ respectively.

Although there is considerable overlap between the distributions
corresponding to the three missions, there are obvious systematic
differences. To illustrate this we combine all the data points with
the use of eqn. 4 (see Fig. 1). The fact that there are points which
obviously fall outside the distribution confirms the presence of
systematic differences between the missions. Indeed the largest
differences occur between \ASCA\ and \ROSAT\ amounting to $\sim 30$
per cent.

Cross-calibration mismatches between missions have been reported by
various authors, in particular between \ROSAT\ and \ASCA\ (e.g.,
Yaqoob et al 1994, Allen \& Fabian 1997). Iwasawa et al (1999)
addressed this point with detail, motivated by simultaneous \ROSAT\
and \ASCA\ observations of the Seyfert galaxy NGC 5548.  They find that
the \ROSAT\ PSPC spectrum is $\Delta\alpha\sim 0.4$ steeper than the
simultaneous \ASCA\ spectrum over a similar  energy band.  Iwasawa et al
(1999) also comment on the fact that these calibration mismatches do
not occur between \ASCA\ and any other missions, which find similar
spectral shapes for a variety of objects observed.

A further complication in the combination of XRB data from various
missions is the different influence of the Galaxy.  The main \ASCA\
(Gendreau et al 1995) and \SAX\ (Vechhi et al 1999) measurements of
the XRB intensity only used events above 1 keV, where the contribution
from the Galaxy is $\sim 1$ per cent or less for these instruments.
However, in the Hasinger (1992) \ROSAT\ estimate, the whole 0.1-2.4
keV band was used to extract both the extragalactic and local
components. For example, more than 30 per cent of the counts above
0.5\, \keV\ are of galactic origin. Although the influence of the
Galaxy on the determination of $I_{XRB}$ is uncertain in the \ROSAT\
data, it is likely to be more important.

Excluding the \ROSAT\ data points, a bayesian estimate of $I_{XRB}$
yields $10.0^{+0.6}_{-0.9}$ $\uint$, but the fact that the
largest measured intensity is the \SAX\ one is suggestive of residual
systematic effects (see again fig. 1). In this case, however, these do
not need to be larger than a few per cent.

%

All of the above estimates give significantly higher values for the
XRB intensity than the extrapolated value of the HEAO-1 data.  For the
\SAX\ and the \ASCA\ data, the {\it HEAO-1} value ($8\, \uint$) is
still marginally consistent with the current modelling in terms of
statistical and cosmic variance of the \ASCA\ and \SAX\ data.
However, this interpretation would mean that the regions of the sky
used in the imaging observations are systematically brighter than the
average sky sampled by {\it HEAO-1}. The possibility that the XRB
spectrum between 1 and 3 keV is steeper than the assumed $\alpha=0.37$
(there is a hint of this in the HEAO-1 data, see the mid panel of
Fig. 4a in Marshall et al 1980) does not really help to alleviate this
mismatch. The reason is that both \ASCA\ and \SAX\ measurements are
quite sensitive to photon energies $\sim 3\, \keV$, where the
intensities found are significantly larger than the {\it HEAO-1}
result at the same energy.

\section{Conclusions}

Cosmic variance is able to account for the dispersion of the measured
values of the extragalactic XRB intensity within the same mission.
However, systematic differences remain among different missions, which
cannot be understood in terms of spatial fluctuations.

A cross-calibration mismatch between \ROSAT\ observations and those
form other missions has been reported (Iwasawa et al 1999).  But even
using the \ASCA\ and \SAX\ data only the result ($10.0^{+0.6}_{-0.9}\,
\uint$) is higher than the XRB intensity at 1 keV extrapolated from
the HEAO-1 data at photon energies above 3 keV.  A steepening of the
XRB spectrum at energies below 3 keV does not really help as both
\ASCA\ and \SAX\ are sensitive to energies $\sim 3\, \keV$ and above
where the mismatch persists. The only conclusion we can reach is that
although internal calibration uncertainties in each mission amount to
probably less than 10 per cent, the cross-calibration among them has
still large residual errors, preventing a more precise determination
of the XRB intensity.

XMM is and will be for many years the most sensitive X-ray imaging
facility over the 0.1-10 keV energy band, providing also moderate
spectral resolution.  Over 1 year, XMM will carry out observations of
500-1000 fields of $\sim 0.2\, \deg^2$, so we expect $\sim 100\,
\deg^2$ to be covered each year.  The cosmic variance (roughly scaling
as $\Omega^{-1/2}$) will then be small enough for unambiguous
determinations of the XRB spectrum and intensity. If the EPIC
instruments can be calibrated to significantly better than 10 per
cent, this will certainly solve the issue of the intensity of the XRB.

\section*{Acknowledgments}

Partial financial support for this work was provided by the
FEDER/CICYT projet 1FD97-0342.

\bsp

\label{lastpage}

\end{document}